\documentstyle[preprint,prd,aps]{revtex}

\advance\hoffset by -3mm  
\advance\voffset by  8mm  

\begin{document}

\bibliographystyle{prsty}

\draft

\author{H.H.~Shih and Y.C.~Lin}

\address{Department of Physics and Astronomy, National Central University,
Chungli, Taiwan 32054}

\title{Renormalization of $K^+ K^- \rightarrow \pi^0\pi^+\pi^-$}
\date{\today}
\maketitle

\narrowtext

\begin{abstract}
We derive the vertices of five-meson and seven-meson anomaly processes
from the four dimensional expansion form of Wess-Zumino term.
Using these vertices we calculate the amplitude, both the finite part
and divergent part, of {$K^+  K^- \rightarrow \pi^0  \pi^+ \pi^-$} to one
loop and renormalize the lagrangian.  The divergent part
agrees with the result derived from path integral approach.
Contribution from counter terms
is estimated by using the vector meson dominance model. Test of the vertex in
the $t$-channel of
$K^- P\rightarrow\Sigma^0\pi^0\pi^+\pi^-$
near the  threshold is discussed. We find that the amplitudes
arising from chiral loop and counter terms are of opposite sign and the
counter term amplitude is about twice the loop amplitude.
\end{abstract}



\section{Introduction}
One of the major successes of theoretical physics in 70's and 80's is the
discovery and the understanding of chiral anomaly\cite{wess71,witt83}. It
reconciles
the discrepancy between the prediction of Sutherland-Veltman theorem and the
experimental data of $\pi^0\rightarrow\gamma\gamma$ and it leads to a deeper
understanding of $PCAC$ theorem. Its connection to differential geometry
opens a door to the studies of anomaly of all sorts.
Experimental results about anomaly are also very fruitful. The data of
$\pi^0\rightarrow\gamma\gamma$ not only successfully prove the existence
of gauged chiral anomaly but also serves as a solid evidence for the fact
that $N_c=3$ in QCD. A few other prototype anomaly processes,
including $\gamma\rightarrow\pi^0\pi^+\pi^-$ and
$\gamma\gamma\rightarrow\pi^0\pi^+\pi^-$, are also extensively
studied\cite{bijn93,bos94} and
proposed in various experiments
\cite{moin94}. The accuracy of these experiments is sensitive to the
higher order corrections to the anomaly processes. In
particular,
it is found that it is necessary to include the $O(p^6)$ contribution in
the momentum power counting in order
to explain the $\gamma\rightarrow\pi^+\pi^-\pi^0$ coupling in the one photon
exchange domain of $\pi^-  Z \rightarrow {\pi^-}^{\prime}  \pi^0
Z^{\prime}$ experiment\cite{moin94}.

It is the purpose of this work to study
the amplitude of pure mesonic anomaly process
$K^+ K^- \rightarrow \pi^0  \pi^+ \pi^-$
to one loop, both the divergent part and the finite part. The former can
be used for the comparison with the divergent part computed from
path integral approach\cite{bijn93,dono89}. It is found that the divergent
part derived from the present approach is the same with the result derived
from path integral approach. The later is useful for the understanding of
higher order
corrections. We also estimate the contribution from the counter terms by using
vector meson dominance model (VMD).
The $K^+ K^- \rightarrow \pi^0  \pi^+ \pi^-$ amplitude can be tested
in the $t$-channel of $K^- P\rightarrow\Sigma^0\pi^0\pi^+\pi^-$ near the
production threshold.
We find that the contribution arising from counter terms and the loop
contribution are of opposite sign. By contraries to the frequently used
working assumption of
chiral perturbation theory, the counter term contribution is twice larger than
the loop contribution. The article is organized
as follows: In Sec. II, we derive the relevant vertices from the four
dimensional form of Wess-Zumino term.
We then calculate
the {$K^+ K^- \rightarrow \pi^0 \pi^+ \pi^-$} amplitude to one loop, both
the divergent part and the finite part.
In Sec. III, we estimate the contribution arising from the counter terms by
using VMD. In Sec. IV, we conclude this paper
by discussing the test of the vertex in the $t$-channel cross
section of $K^- P\rightarrow\Sigma^0\pi^0\pi^+\pi^-$.

\section{Loop Amplitude}
The closed form of Wess-Zumino terms only exists
in five-dimensional manifold\cite{witt83} and the form is given by
\begin{eqnarray}
\Gamma & = & \int_Q \omega_{ijklm} d\Sigma^{ijklm}\nonumber\\
       & = & -{iN_c \over 240 \pi^2} \int_Q d\Sigma^{ijklm} \left[
             TrU^{-1}{\partial U\over \partial y^i}
               U^{-1}{\partial U\over \partial y^j}
               U^{-1}{\partial U\over \partial y^k}
               U^{-1}{\partial U\over \partial y^l}
               U^{-1}{\partial U\over \partial y^m}
               \right],
\label{eq1}
\end{eqnarray}
where $d\Sigma^{ijklm}=d^5y\epsilon^{ijklm}=dy^i\wedge dy^j\wedge dy^k\wedge
dy^l\wedge dy^m$ is the five-form defined on the five-dimensional manifold
$Q$ and
\begin{eqnarray}
U & = & \exp {i \over f_{\pi}} \phi \nonumber\\
  & = & 1 + {i\over f_{\pi}}\phi + {1 \over 2} {\left( i\over
        f_{\pi}\right)}^2\phi^2 + \cdots
\label{eq2}
\end{eqnarray}
is the nonlinear realization of meson octet $\phi$, $\phi= \phi^a\lambda_a$.
The four dimension form can be derived by first expanding $U$ in terms of
$\phi$ and then partially
integrating the five dimensional form. The relevant pieces to the loop
amplitude calculation of
$K^+K^-\pi^0\pi^+\pi^-$ are the five-meson coupling
\begin{eqnarray}
\Gamma & = & {N_c \over 240 \pi^2 f_{\pi}^5} \int d^5 x
             \epsilon^{ijklm}
             (\partial_i\phi \partial_j\phi
             \partial_k\phi \partial_l\phi\partial_m\phi)\nonumber\\
       & = & {N_c \over 240 \pi^2 f_{\pi}^5} \int d^5 x
             \epsilon^{ijklm}
             \partial_i(\phi \partial_j\phi
             \partial_k\phi \partial_l\phi\partial_m\phi)\nonumber\\
       & = & {N_c \over 240 \pi^2 f_{\pi}^5} \int d^4 x
             \epsilon^{\mu\nu\alpha\beta}
             \phi\partial_{\mu}\phi \partial_{\nu}\phi
             \partial_{\alpha}\phi \partial_{\beta}\phi ,
\label{eq3}
\end{eqnarray}
which gives rise to the tree level amplitude of
{$K^+  K^- \rightarrow \pi^0 \pi^+ \pi^-$}
\begin{eqnarray}
M(K^+(P_4)K^-(P_5)\rightarrow\pi^0(P_1)\pi^+(P_2)\pi^-(P_3)) & = & {i3 \over
4\pi^2 f_{\pi}^5}
\epsilon^{\mu\nu\alpha\beta}k_{2\mu}k_{3\nu}k_{4\alpha}k_{5\beta},
\label{eq4}
\end{eqnarray}
and the seven-meson coupling
\begin{eqnarray}
\Gamma & = & {- N_c\over 4032 f_{\pi}^7} \int d\Sigma^{ijklm}Tr[
8\partial_i\phi^3 \partial_j\phi \partial_k\phi \partial_l\phi \partial_m\phi
\nonumber\\
& & -9 \partial_i\phi^2 \partial_j\phi^2 \partial_k\phi \partial_l\phi
\partial_m\phi
+3\partial_i\phi^2 \partial_j\phi \partial_k\phi^2 \partial_l\phi
\partial_m\phi ]\nonumber\\
& = & {- N_c\over 4032 f_{\pi}^7} \int d^4x\epsilon^{\mu\nu\alpha\beta} Tr[
8\phi^3 \partial_{\mu}\phi \partial_{\nu}\phi \partial_{\alpha}\phi
\partial_{\beta}\phi\nonumber\\
& & -9\phi^2 \partial_{\mu}\phi^2 \partial_{\nu}\phi \partial_{\alpha}\phi
\partial_{\beta}\phi
+3\phi^2 \partial_{\mu}\phi \partial_{\nu}\phi^2 \partial_{\alpha}\phi
\partial_{\beta}\phi ].
\label{eq5}
\end{eqnarray}
We listed the relevant four-meson vertices arising from the normal parity part
of chiral
lagrangian (see, for example, $ref.$ \cite{gass85}) in Table I.
The five-meson vertices and seven-meson vertices derived from $eq.$ 4 and $eq.$
5 are listed in Table II and Table III respectively.

The loop amplitude receives three kinds of contribution
: wave function renormalization (Fig. 1), meson
decay constant renormalization (Fig. 2) and chiral loop contribution
(Fig. 3). All these contributions contain a
divergent part and a finite part.
The divergent part in loop amplitude has to be cancelled by the
divergent part in counter terms of higher momentum
power and it is in general not interesting to the phenomenological concern.
However, as is
pointed out in \cite{dono89}, the fact that the divergent part only exists
in four dimensional manifold implies
the non-renormalization of Wess-Zumino-Witten anomaly. So it worths
calculating
the divergent part by the present approach and compare it with the result
derived from path integral approach.
The finite part usually contains
terms involving nontrivial momentum dependence which can not be
absorbed into the counter terms. This part, together with the finite part
in the counter terms, are of interest to phenomenological concern.

Both the results of meson decay constant
renormalization and wave function renormalization are well-known\cite{gass85},
and we only quote the results here for the sake of convenience:
\begin{eqnarray}
f_{\pi} & = & f(1+A_{m_{\pi}}+{1\over 2}A_{m_{K}})\nonumber\\
f_K & = & f(1+{3\over 8}A_{m_{\pi}}+{3\over 4}A_{m_K}+{3\over 8}A_{m_{\eta}})
\label{eq6}
\end{eqnarray}
and
\begin{eqnarray}
Z_{\pi} & = & 1+{2\over 3}A_{m_{\pi}}+{1\over 3}A_{m_K}\nonumber\\
Z_K & = & 1+{1\over 4}A_{m_{\pi}}+{1\over 2}A_{m_{K}}+{1\over 4}A_{m_{\eta}},
\label{eq7}
\end{eqnarray}
where
\begin{eqnarray}
A_m & = & {1\over 16\pi^2 f^2}m^2\lambda-{1\over 16\pi^2 f^2}m^2\ln m^2
\nonumber\\
\lambda & = & {2\over \epsilon} - \gamma_E + 1 +\ln 4\pi,
\end{eqnarray}
in which $\epsilon = 4-d$ and $\gamma_E$ is the Euler number.
There are no diagrams correspond
to the meson decay constant renormalization in the loop calculation. The
reason why we include these
contribution is that we use the physical meson decay constants $f_{\pi}$ and
$f_K$ in the final expression instead of using the bare one $f$, and the
contributions in $eq.$ 6 account for the difference.

The loop integrals which are relevant to our
calculations are
listed in Appendix. The overall results are
summarized in the following equation:
\begin{eqnarray}
A_{one-loop}& = & i{ 3 \over {4 \pi^2 f^5}} \epsilon^{\mu \nu \alpha \beta}
P_{2\mu}P_{3\nu}P_{4\alpha}P_{5\beta} \{ {1 \over {16 \pi^2 f^2}}
\nonumber\\
&   & [ {1 \over 12} \lambda (3 P_{45}^2+2 P_{12}^2+2 P_{23}^2+2 P_{13}^2+
2 P_{25}^2+2 P_{34}^2+P_{14}^2+P_{15}^2) +
\nonumber\\
&   & {1 \over 4}F(P_{14}^2,P_{14}^2+M_{\pi}^2-M_K^2,M_{\pi}^2)+
{1 \over 4} F(P_{14}^2,P_{14}^2+M_{\eta}^2-M_K^2,M_{\eta}^2)+
\nonumber\\
&  & {1 \over 4}F(P_{15}^2,P_{15}^2+M_{\pi}^2-M_K^2,M_{\pi}^2)+
{1 \over 4}F(P_{15}^2,P_{15}^2+M_{\eta}^2-M_K^2,M_{\eta}^2)+
\nonumber\\
&  & {1 \over 2}F(P_{25}^2,P_{25}^2+M_{\pi}^2-M_K^2,M_{\pi}^2)+
{1 \over 2}F(P_{25}^2,P_{25}^2+M_{\eta}^2-M_K^2,M_{\eta}^2)+
\nonumber\\
&  & {1 \over 2}F(P_{34}^2,P_{34}^2+M_{\pi}^2-M_K^2,M_{\pi}^2)+
{1 \over 2}F(P_{34}^2,P_{34}^2+M_{\eta}^2-M_K^2,M_{\eta}^2)+
\nonumber\\
&  & {3 \over 2}F(P_{45}^2,P_{45}^2,M_K^2)+F(P_{12}^2,P_{12}^2,
M_{\pi}^2) +F(P_{23}^2,P_{23}^2,M_{\pi}^2)+F(P_{13}^2,P_{13}^2,
M_{\pi}^2)-
\nonumber\\
&  & {{15 \over 4}} M_{\pi}^2 ln {M_{\pi}^2\over \Lambda^2} - 3 M_{K}^2 ln
{M_{K}^2\over \Lambda^2} - {3 \over 4} M_{\eta}^2 ln {M_{\eta}^2\over
\Lambda^2} ] \} \nonumber\\
\end{eqnarray}
where
$P_{ij}^2\equiv
(P_i + P_j)^2$, $\Lambda=4\pi f_{\pi}\sim 1$ $GeV$ and the function
$F(p^2,q^2,m^2)$
is the result of loop integration which is defined in Appendix .
The divergent part, which is proportional to $\lambda$, is in agreement with
the result in \cite{bijn93} but differs with that in \cite{dono89} by a
factor of $1/2$. In any case, it also leads to the same result that the
divergent terms only exist in four dimensional manifold.
The situation for the finite part is somewhat different from the case of
$\pi^0\rightarrow\gamma\gamma$.
In $\pi^0\rightarrow\gamma\gamma$, the finite part of chiral corrections
combine in such a way that the net effect of higher order correction is just to
replace the bare pion decay constant $f$ in tree level amplitude by the
physical one $f_{\pi}$.
This is also the realization of
the non-renormalization of chiral anomaly\cite{dono85}. This fact is not
manifest in the present
case. The reason is mainly due to the nontrivial dependence of momentum. Even
at energy just above the threshold of $K^+K^-$, the pions have very large
$P_{ij}^2$ and this makes the cancellation non-trivial.

\section{Counter Terms}
The forms of the counter terms in chiral lagrangian are dictated by
symmetries. However, the coefficients of counter terms are determined by the
detailed underlying dynamics of QCD at energy higher than the chiral cutoff
scale and they remain as unknowns in any practical use of chiral lagrangian.
The value of these coefficients can be either fitted by experiments or
predicted by models. Since the number of existing data in the anomaly sector
is not enough
to decide all the coefficients, we are forced to resort to model prediction.
VMD is the most natural choice of model due to the fact that most
of the processes receiving contribution from W.Z.W. are radiative processes.

Adopting the hidden symmetry approach\cite{band88}
and integrating out the $\rho$ meson which is considered
as heavy particle\cite{bijn93} in the process of interest, the relevant part
of anomaly lagrangian is given by
\begin{eqnarray}
{\it L}_{anomaly}^{6} &=& {3 \over 8\sqrt{2}\pi^2 f^5_{\pi}} \epsilon^
{\mu \nu \alpha \beta} {1 \over M_{\rho}^2 } {\sl tr} \{ \partial_{\mu}
\phi \partial_{\nu}\phi \partial_{\alpha}\phi [ \partial_{\lambda}
\partial_{\beta}\phi \partial^{\lambda}\phi \nonumber\\
&& + \partial_{\beta}\phi \partial^2
\phi - \partial^2 \phi \partial_\beta \phi - \partial_{\lambda}\phi \partial_
{\beta} \partial^{\lambda}\phi ] \},
\end{eqnarray}
where we have assumed complete vector meson dominance in deriving the
coefficients. The amplitude then reads
\begin{eqnarray}
A_{counter\: term} &=& {\it i}{3 \over 4 \pi^2
f_{\pi}^5 } \epsilon^{\mu \nu \alpha \beta} P_{2 \mu}P_{3 \nu}P_{4 \alpha}
P_{5 \beta} \{ {1 \over 4M_{\rho}^2} (3 P_{45}^2 +2 P_{34}^2+2 P_{25}^2
\nonumber\\
&& + 2 P_{12}^2 +2 P_{13}^2+2 P_{23}^2 +P_{14}^2 +P_{15}^2 ) \}.
\end{eqnarray}
Note that the tensor structure of the counter term amplitude is
identical to that of one-loop amplitude, as it is expected. Comparison
between the size of
loop amplitude and the size of counter term amplitude requires evaluation of
amplitude at a specified kinetics, we will leave the comparison to Sec. IV.

\section{Discussion and Conclusion}
The $K^+ K^- \rightarrow \pi^0\pi^+\pi^-$ amplitude can be tested in
the $t$-channel
of meson-baryon scattering. The scattering process $K^- P\rightarrow
\Sigma^0 \pi^0\pi^+\pi^-$ (Fig. 4) is the most favorable process for the
study of the $K^+ K^- \rightarrow \pi^0\pi^+\pi^-$ amplitude for both the
experimental and theoretical
considerations. On the experimental side, this process is the by-product of
polarized hyperon production, e.g. $E811$ at Brookhaven AGS\cite{lars93}.
The mode $K^- P\rightarrow\Sigma^0\pi^0$ has been carefully studied. The study
of the mode $K^- P\rightarrow \Sigma^0 \pi^0\pi^+\pi^-$ requires only
additional detection of $\pi^+$ and $\pi^-$. On the theoretical side, we
like to keep all the particles involved carrying momentum as low as possible so
that we have
better control of the momentum expansion perturbation. Keeping a kaon in the
internal propagator allows us to minimize the momentum carried by
the rest particles. Given this most favorable kinetic situation, however, the
energy range which the prediction is reliable is very limited. The denominator
of momentum expansion in chiral perturbation theory is believed to be $4\pi
f_{\pi}\sim$ 1 $GeV$. The mass of proton, though is close to 1 GeV, is not
involved in the chiral dynamics and will not cause any problem in momentum
expansion. But the minimum total energy of kaon required to produce the final
state particles $E=
M_{\Sigma} - M_P + 3 m_{\pi}\sim$ 750 $MeV$ is large and only the prediction
near the production threshold is trustworthy.

The $t$-channel amplitude can be calculated by baryon chiral
perturbation theory\cite{jenk91}. We adopt the value of $D=0.8$ and $F=0.5$
coefficients fitted by tree level hyperon noleptonic decays\cite{jaff90}.

In Fig. 5 and Fig. 6 we plot the $t$-channel differential cross section with
repect to $E^*_{\Sigma}$ and $S_2= (P_{\pi^+} + P_{\pi^-})^2$
correspondingly. The $t$-channel cross
section of various combinations of amplitudes is depicted in Fig. 7. We find
that the amplitude of counter terms is of the same sign as the tree level
amplitude but opposite to the loop amplitude. The occurrence of cancellation
between loop amplitude and counter term amplitude makes the correction of cross
section at $30\%$ level near the threshold, which is better than what is
expected
in this energy range.  The size of counter term amplitude is about twice
than that of loop amplitude. This is somewhat unexpected. In chiral
perturbation theory, dominance of loop amplitude over the counter term
amplitude is a frequently used working assumption.

In summary,
we calculate the amplitude of the prototype pure mesonic
anomaly
$K^+ K^-\rightarrow\pi^0\pi^+\pi^-$ to one loop, both the divergent part and
the finite part. The former agrees with the derivation by path integral
approach, which confirms again the non-renormalization of anomaly.
The counter term contribution is estimated
by using VMD. The $K^+ K^-\rightarrow\pi^0\pi^+\pi^-$ amplitude can be tested
in the $t$-channel of $K^-
P\rightarrow \Sigma^0 \pi^0\pi^+\pi^-$ experiment. The counter term
contribution and loop contribution add up to $30\%$ correction in cross
section to the tree level amplitude near the production threshold.

\acknowledgments

This research is supported in part by National Science Council in Taiwan
under Contract No. NSC84-2122-M008-013 and NSC-2732-M008-001. The
authors like to thank S.\ C.\ Lee  and Y. H. Chang for useful conversation.

{\section*{Appendix:}}
The integrals required in the calculation of loop amplitude are given below.
The functions $I(r,m)$ and $I_{\mu \nu}(m)$ are used in the intermediate steps
of calculation. Their definitions and results are given by
\begin{eqnarray}
I(r,m) & \equiv & \int {{d^n l}\over (2\pi )^n} {(l^2)^r \over (l^2-R^2)^m}
\nonumber\\
&    =   & i{ (-1)^{r-m}\over (16\pi ^2)^{n/4}}  {(R^2)^{r-m+n/2}}
{{\Gamma (n/2+r)\Gamma (m-r-n/2)} \over { \Gamma (n/2) \Gamma (m) }}
\nonumber
\end{eqnarray}
and
\begin{eqnarray}
I_{\mu \nu}(m) & \equiv & \int {{d^n l}\over  (2\pi)^n} {{l_{\mu} l_{\nu}}
\over {(l^2-R^2)^m}}\nonumber\\
&    =   & {1 \over n} {g_{\mu \nu} I (1,m)}.
\nonumber
\end{eqnarray}
The function $F(p^2,q^2,m^2)$ is used to express the final result of chiral
loop corrections. The definition and the result on integral are given by
\begin{eqnarray}
F(p^2,q^2,m^2) & \equiv & \int_{0}^{1} dz (p^2 z^2 -q^2 z +m^2 )
\ln {(p^2 z^2-  q^2 z + m^2 )\over \Lambda^2}\nonumber\\
&    =   & ({1 \over 3} - {1 \over 2} y +x) p^2 \ln {( p^2 - q^2
+ m^2)\over \Lambda^2} - ( {2 \over 9} - {1 \over 3} y - {1 \over 6} y^2 +{4
\over 3} x ) p^2 \nonumber\\
&        & + ( {1 \over 12} y^2 - {1 \over 3} x ) \sqrt{ y^2 - 4x } p^2
\ln {{2x -y +\sqrt { y^2 -4x}} \over {2x-y- \sqrt {y^2-4x}}}\nonumber\\
&        & - ( {1 \over 2} xy - {1 \over 12} y^3) p^2 \ln {{x-y+1} \over x},
\nonumber
\end{eqnarray}
where
\begin{eqnarray*}
x \equiv  {m^2 \over p^2 },
y \equiv  {q^2 \over p^2 }.
\nonumber
\end{eqnarray*}

\newpage

\figure{{\bf Fig 1:} Wave function renormalizations of $\pi$ and $K$.
\label{fig1}}

\figure{{\bf Fig. 2:} Decay constant renormalizations of $f_{\pi}$ and
$f_K$.
\label{fig2}}

\figure{{\bf Fig. 3:} Contribution arising from the genuine chiral loop to
$K^+  K^- \rightarrow \pi^0  \pi^+ \pi^-$.
\label{fig3}}

\figure{{\bf Fig. 4:} The scattering process $K^- P\rightarrow
\Sigma^{0}\pi^0\pi^+\pi^-$
which can be used to probe the $K^+ K^- \rightarrow \pi^0  \pi^+ \pi^-$ vertex.
\label{fig4}}

\figure{{\bf Fig. 5:} The differential cross section ${d\sigma\over
dE^*_{\Sigma}}$ ($E^*_{\Sigma}$ is the energy of $\Sigma^0$ in the center of
mass frame) of the $t$-channel amplitude in unit of
$10^{-2}nb/GeV^2$ near the production threshold, $S=(P_{K^-} + P_P)^2$.
The dashed line is the tree diagram contribution only while the solid line is
the total contribution including tree diagrams, chiral loops and counter terms.

\figure{{\bf Fig. 6:} The differential cross section ${d\sigma\over ds_2}$
($s_2= (\pi^+ +\pi^-)^2$) of the $t$-channel amplitude in unit of
$10^{-2}nb/GeV^2$ near the production threshold.
The dashed line is the tree diagram contribution only while the solid line is
the total contribution including tree diagrams, chiral loops and counter terms.

\figure{{\bf Fig. 7:} The $t$-channel cross section (in unit of $10^{-6}nb$) of
various
combinations of contributions ($E_{CMS}=\sqrt{s}$). (a) tree + counter
terms (b) tree + counter terms + loop (c) tree (d) tree + loop (e) counter
terms (f) loop.

 \newpage
\begin{table}
\caption{{\bf Four-meson Vertices:}}
\begin{tabular}{lll}

${\pi}^o(P_1){\pi}^o(P_2){\pi}^+(P_3){\pi}^-(P_4)$ &
${i \over {3 f^2}}[2(P_1P_2+P_3P_4)-(P_1+P_2)(P_3+P_4)+M_{\pi}^2]$ \\

${\pi}^+(P_1){\pi}^+(P_2){\pi}^-(P_3){\pi}^-(P_4)$ &
${i \over {3 f^2}}[(P_1+P_3)(P_2+P_4)+(P_1+P_4)(P_2+P_3)$ \\
 & $ -4(P_1 P_2+P_3 P_4)+2M_{\pi}^2]$ \\

${K}^+(P_1){K}^+(P_2){K}^-(P_3){K}^-(P_4)$ &
${i \over {3 f^2}}[(P_1+P_3)(P_2+P_4)+(P_1+P_4)(P_2+P_3)$ \\
 & $ -4(P_1 P_2+P_3 P_4)+2M_{K}^2]$ \\

${K}^+(P_1){K}^-(P_2){K}^o(P_3)\bar{K}^o(P_4)$ &
${i \over {6 f^2}}[(P_1+P_2)(P_3+P_4)+(P_1+P_4)(P_2+P_3)$ \\
 & $-4(P_1 P_3+P_2 P_4)+2M_{K}^2]$ \\

${\pi}^o(P_1){\pi}^o(P_2){K}^+(P_3){K}^-(P_4)$ &
$-{i \over {12 f^2}}[(P_1+P_2)(P_3+P_4)-2(P_1P_2+P_3P_4)$ \\
 & $-2(M_{\pi}^2+M_{K}^2)]$ \\

${\pi}^+(P_1){\pi}^-(P_2){K}^+(P_3){K}^-(P_4)$ &
${i \over {6 f^2}}[(P_1+P_2)(P_3+P_4)+(P_1+P_4)(P_2+P_3)$ \\
 & $-4(P_1 P_3+P_2 P_4)+(M_{K}^2+M_{\pi}^2)]$ \\

${\pi}^+(P_1){\pi}^-(P_2){K}^o(P_3)\bar{K}^o(P_4)$ &
${i \over {6 f^2}}[(P_1+P_2)(P_3+P_4)+(P_1+P_3)(P_2+P_4)$ \\
 & $-4(P_1 P_4+P_2 P_3)+(M_{K}^2+M_{\pi}^2)]$ \\

${K}^o(P_1)\bar{K}^o(P_2){K}^o(P_3)\bar{K}^o(P_4)$ &
${i \over {3 f^2}}[(P_1+P_2)(P_3+P_4)+(P_1+P_4)(P_2+P_3)$ \\
 & $-4(P_1 P_3+P_2 P_4)+2M_{K}^2]$\\

${\pi}^o(P_1){\eta}(P_2){K}^+(P_3){K}^-(P_4)$ &
${i \over {4 \sqrt 3 f^2}}[2(P_1+P_2)^2-(P_1+P_3)^2-(P_1+P_4)^2$\\
 & $+{2 \over 3}(M_\pi^2-M_K^2)]$ \\

${\pi}^+(P_1){K}^-(P_2){K}^o(P_3){\eta}(P_4)$ &
${i \over {6 \sqrt 6 f^2}}M_{\pi}^2({m_u-m_s \over m_u}) -
 {i \over {2 \sqrt 6 f^2}}[P_1P_2+P_1P_3$ \\
 & $+P_2P_4+P_3P_4-2 P_1P_4-2 P_2P_3]$ \\

${\pi}^-(P_1){K}^+(P_2)\bar{K}^o(P_3){\eta}(P_4)$ &
${i \over {6 \sqrt 6 f^2}}M_{\pi}^2({m_u-m_s \over m_u}) -
 {i \over {2 \sqrt 6 f^2}}[P_1P_2+P_1P_3$ \\
 & $+P_2P_4+P_3P_4-2 P_1P_4-2 P_2P_3]$ \\

\end{tabular}
\end{table}

\begin{table}
\caption{{\bf Five-meson Vertices:}}
\begin{tabular}{lll}

${\pi}^0(P_1){\pi}^+(P_2){\pi}^-(P_3){K}^+(P_4){K}^-(P_5)$ &
${i3 \over {4 \pi^2 f^5}}\epsilon^{\mu\nu\alpha\beta}
 P_{2\mu}P_{3\nu}P_{4\alpha}P_{5\beta}$ \\

${\pi}^0(P_1){\pi}^+(P_2){\pi}^-(P_3){K}^o(P_4)\bar{K}^o(P_5)$ &
${i 3 \over {4 \pi^2 f^5}}\epsilon^{\mu\nu\alpha\beta}
 P_{2\mu}P_{3\nu}P_{4\alpha}P_{5\beta}$ \\

${\eta}(P_1){\pi}^+(P_2){\pi}^-(P_3){K}^+(P_4){K}^-(P_5)$ &
${i \sqrt 3 \over {4 \pi^2 f^5}}\epsilon^{\mu\nu\alpha\beta}
 P_{2\mu}P_{3\nu}P_{4\alpha}P_{5\beta}$ \\

${\pi}^0(P_1){\pi}^+(P_2){K}^-(P_3){K}^o(P_4){\eta}(P_5)$ &
$-{i 3 \over {2 \sqrt 6 \pi^2 f^5}}\epsilon^{\mu\nu\alpha\beta}
 P_{2\mu}P_{3\nu}P_{4\alpha}P_{5\beta}$ \\

${\pi}^0(P_1){\pi}^-(P_2){K}^+(P_3)\bar{K}^o(P_4){\eta}(P_5)$ &
$-{i 3 \over {2 \sqrt 6 \pi^2 f^5}}\epsilon^{\mu\nu\alpha\beta}
 P_{2\mu}P_{3\nu}P_{4\alpha}P_{5\beta}$ \\
\end{tabular}
\end{table}

\begin{table}
\caption{{\bf Seven-meson Vertices:} Any piece in the vertices involving
the internal loop momentum vanishes after evaluating the loop integral. The
remaining part which contributes to the loop amplitude involves external
momentum only.}

\begin{tabular}{lll}

${\pi}^0(P_1){\pi}^+(P_2){\pi}^-(P_3){K}^+(P_4){K}^-(P_5){K}^+(P_6){K}^-(P_7)
$ &
$-{i \over {2 \pi^2 f^7}}\epsilon^{\mu\nu\alpha\beta}
 P_{2\mu}P_{3\nu}P_{4\alpha}P_{5\beta}$ \\

${\pi}^0(P_1){\pi}^+(P_2){\pi}^-(P_3){K}^+(P_4){K}^-(P_5){K}^o(P_6)
 \bar{K}^o(P_7)$ &
$-{i \over {4 \pi^2 f^7}}\epsilon^{\mu\nu\alpha\beta}
 P_{2\mu}P_{3\nu}P_{4\alpha}P_{5\beta}$ \\

${\pi}^0(P_1){\pi}^+(P_2){\pi}^-(P_3){K}^+(P_4){K}^-(P_5){\pi}^+(P_6){\pi}^-
(P_7) $ & $-{i 5 \over {8 \pi^2 f^7}}\epsilon^{\mu\nu\alpha\beta}
 P_{2\mu}P_{3\nu}P_{4\alpha}P_{5\beta}$ \\

${\pi}^0(P_1){\pi}^+(P_2){\pi}^-(P_3){K}^+(P_4){K}^-(P_5){\pi}^0(P_6){\pi}^0
(P_7) $ & $-{i 5\over {16 \pi^2 f^7}}\epsilon^{\mu\nu\alpha\beta}
 P_{2\mu}P_{3\nu}P_{4\alpha}P_{5\beta}$ \\

${\pi}^0(P_1){\pi}^+(P_2){\pi}^-(P_3){K}^+(P_4){K}^-(P_5){\eta}(P_6){\eta}
(P_7) $ & $-{i 3\over {8 \pi^2 f^7}}\epsilon^{\mu\nu\alpha\beta}
 P_{2\mu}P_{3\nu}P_{4\alpha}P_{5\beta}$ \\

\end{tabular}
\end{table}
\end{document}